\documentclass[preprintnumbers,prd,floatfix,twocolumn,amsmath,amssymb,nofootinbib]{revtex4}

\usepackage{latexsym}
\usepackage{amssymb}
\usepackage{amsmath}
\usepackage{dcolumn}
\usepackage{bm}
\usepackage{comment}
\usepackage{color}

\begin{document}

\title{On phenomenological models of dark energy interacting with dark matter}
\author{Nicola Tamanini}
\email{nicola.tamanini@cea.fr}
\affiliation{Institut de Physique Th{\'e}orique, CEA-Saclay, F-91191, Gif-sur-Yvette, 
France}

\begin{abstract}
An interaction between dark matter and dark energy is usually introduced by a phenomenological modification of the matter conservation equations, while the Einstein equations are left unchanged.
Starting from some general and fundamental considerations, in this work it is shown that a coupling in the dark sector is likely to introduce new terms also in the gravitational dynamics.
Specifically in the cosmological background equations a bulk dissipative pressure, characterizing viscous effects and able to suppress structure formation at small scales, should appear from the dark coupling.
At the level of the perturbations the analysis presented in this work reveals instead the difficulties in properly defining the dark sector interaction from a phenomenological perspective.
\end{abstract}

\maketitle

\section{Introduction}

A truly satisfactory explanation for the observed accelerated expansion of the universe still evades our understanding of Nature.
More than fifteen years have passed since this acceleration was first discovered by type-Ia supernovae surveys \cite{Riess:1998cb,Perlmutter:1998np}, and by now the need for {\it dark energy} (DE), an electromagnetically transparent cosmological source accounting for such phenomenon, is widely accepted by the scientific community \cite{Planck:2015xua}.
Similarly astronomical data at galactic and galaxy clusters distances have accumulated evidence for the existence of mass beyond the luminous matter directly observed by telescopes, namely {\it dark matter} (DM).
Such unobservable mass distribution has been made responsible for the anomalous rotational motion of galaxies as well as for influencing the dynamical growth of cosmic structures \cite{Clowe:2006eq}.
Both DE and DM can be detected only through the gravitational effects they induce on the directly observable matter, constituted by Standard Model particles.

The most popular cosmological theories postulate that no interaction between the two dark components is present beyond their mutual gravitational attraction.
These include the standard $\Lambda$CDM model, where DM is provided by a massive non-relativistic particle and DE is nothing but the cosmological constant.
Nevertheless, although no coupling with baryonic matter is allowed by the observations, nothing prevents DM and DE to interact by some non-gravitational mechanism, possibly exchanging energy and momentum between them.
Of course in such theories DE is required to be a dynamical quantity which is usually associated either to some fundamental fields (particles) or to some modifications of gravity \cite{2010deto.book.....A}, although a quantum running of the cosmological constant could still in principle provide an evolving scenario; see e.g.~\cite{Sola:2007sv,Shapiro:2009dh,Sola:2013gha}.
In what follows DE will be generally described by a cosmological fluid dynamically connected to some unspecified fundamental degrees of freedom which could equally be of particle or gravitational origin.

At the background cosmological level, with the universe assumed to be perfectly homogeneous and isotropic, the standard phenomenological approach to coupling DE to DM consists in allowing for an exchange of energy between the two matter conservation equations 
\begin{align}
	\dot{\rho}_{dm} + 3H (\rho_{dm} + p_{dm}) &= -Q \,, \label{eq:matter_dm} \\
	\dot{\rho}_{de} + 3H (\rho_{de}+ p_{de}) &= Q \,, \label{eq:matter_de}
\end{align}
where a dot denotes differentiation with respect to coordinate time, $H$ is the Hubble rate, $\rho_{dm}$ and $\rho_{de}$ are the energy densities, $p_{dm}$ and $p_{de}$ are the pressures and $Q$ characterizes the amount of energy exchanged.
If $Q>0$ DM releases energy into DE, while for $Q<0$ the energy flows in the opposite direction.
The coupling is introduced in order to not spoil the total energy conservation
\begin{equation}
	\dot{\rho}_{\rm tot} + 3H (\rho_{\rm tot} + {p}_{\rm tot}) = 0 \,,
\end{equation}
where $\rho_{\rm tot} = \rho_{dm} + \rho_{de}$ and $p_{\rm tot} = p_{dm} + p_{de}$.
Ignoring the contribution of baryons and radiation, Eqs.~(\ref{eq:matter_dm}) and (\ref{eq:matter_de}) are completed by the standard cosmological equations
\begin{align}
	3H^2 &= \rho_{dm} + \rho_{de} \,, \label{eq:Friedmann} \\
	2\dot{H} + 3H^2 &= -p_{dm} -p_{de} \,, \label{eq:acceleration} 
\end{align}
and by the two equations of state (EoS)
\begin{equation}
	p_{dm}=w_{dm}\rho_{dm} \quad\mbox{and}\quad p_{de}=w_{de}\rho_{de} \,,
	\label{eq:EoS_linear}
\end{equation}
where $w_{dm}$ and $w_{de}$ are EoS parameters with common ($\Lambda$CDM) values of $w_{dm}=0$ and $w_{de}=-1$.
Note that in Eqs.~(\ref{eq:Friedmann})--(\ref{eq:acceleration}) spatial flatness has been assumed.
However the analysis presented in this work can equally be repeated when non-zero curvature is taken into account.

DE models interacting with DM have been extensively studied in several works.
They are well known for providing a possible solution to the cosmic coincidence problem \cite{Chimento:2003iea}, as well as yielding a physical explanation for measuring a DE phantom EoS \cite{Das:2005yj}.
Moreover an increasing amount of observational signatures for an interaction in the dark sector has been recently pointed out in several works, e.g.~\cite{Salvatelli:2013wra,Salvatelli:2014zta,Li:2013bya,Li:2014cee,Abdalla:2014cla,Ade:2015rim,Valiviita:2015dfa}.
Although at the moment this evidence is not statistically robust, future probes are expected to provide more accurate constraints \cite{Geng:2015ara,Giocoli:2015tka,Duniya:2015nva,Yin:2015pqa}.

In the absence of a fundamental, microscopical description of the dark sector, all such models are necessarily phenomenological.
In particular the coupling $Q$ appearing in Eqs.~(\ref{eq:matter_dm}) and (\ref{eq:matter_de}), and characterizing the interaction at the background level, can only be taken arbitrarily.
The most popular interacting terms considered in the literature are $Q\propto \rho$ and $Q\propto H \rho$ where $\rho$ stands for either the DE or DM energy density or even a combination of the two.
For an outline of the literature on the subject the reader might refer to \cite{Valiviita:2015dfa} and the detailed references therein.
In each one of these phenomenological models Eqs.~(\ref{eq:matter_dm})--(\ref{eq:EoS_linear}) are employed to characterize the background cosmological dynamics.
But do they represent the most generally possible equations for such models?
Or in other words: can the DE-DM coupling appear somehow differently at the background level?

\section{Background Equations}

To address this question Eqs.~(\ref{eq:Friedmann})--(\ref{eq:EoS_linear}) will be derived from a more fundamental approach.
The cosmological dynamics of both DE and DM will be assumed to be provided by some hypothetical Lagrangian written as
\begin{equation}
	\mathcal{L}_{\rm tot} = \mathcal{L}_{\rm GR} + \mathcal{L}_{\rm dm}(\psi_{dm}) + \mathcal{L}_{\rm de}(\psi_{de}) + \mathcal{L}_{\rm int}(\psi_{dm},\psi_{de}) \,,
	\label{eq:Lag_tot}
\end{equation}
where $\mathcal{L}_{\rm GR}$ stands for the standard Einstein-Hilbert Lagrangian and $\psi_{dm}$ and $\psi_{de}$ collectively denote the unknown degrees of freedom of DM and DE.
The ``bare'' Lagrangians of DM and DE are given by $\mathcal{L}_{\rm dm}(\psi_{dm})$ and $\mathcal{L}_{\rm de}(\psi_{de})$, while all interacting terms, where the DE variables couple to the DM ones, have been collected into $\mathcal{L}_{\rm int}(\psi_{dm},\psi_{de})$.
The cosmological background equations following from the variation of the Lagrangian (\ref{eq:Lag_tot}) can be written as
\begin{gather}
	3H^2 = \tilde\rho_{dm} + \tilde\rho_{de} + \tilde\rho_{\rm int} \,, \label{eq:Friedmann_tilde} \\
	2\dot{H} + 3H^2 = -\tilde{p}_{dm} -\tilde{p}_{de} -\tilde{p}_{\rm int} \,, \label{eq:acceleration_tilde} \\
	\dot{\tilde\rho}_{dm} + 3H (\tilde\rho_{dm} + \tilde{p}_{dm}) = \tilde{Q}_{dm} \,, \label{eq:matter_tilde_dm} \\
	\dot{\tilde\rho}_{de} + 3H (\tilde\rho_{de} + \tilde{p}_{de}) = \tilde{Q}_{de} \,, \label{eq:matter_tilde_de}
\end{gather}
where the tilde notation is employed to distinguish quantities derived from the variation of the Lagrangian (\ref{eq:Lag_tot}) with the corresponding counterparts in Eqs.~(\ref{eq:matter_dm})--(\ref{eq:EoS_linear}).
In general $\tilde{Q}_{dm}$ and $\tilde{Q}_{dm}$ are related to the variation of $\mathcal{L}_{\rm int}(\psi_{dm},\psi_{de})$ with respect to $\psi_{dm}$ and $\psi_{de}$, while $\tilde\rho_{\rm int}$ and $\tilde{p}_{\rm int}$ are derived from the variation of $\mathcal{L}_{\rm int}(\psi_{dm},\psi_{de})$ with respect to the metric.

One can immediately notice that Eqs.~(\ref{eq:Friedmann_tilde})--(\ref{eq:matter_tilde_de}) differ from Eqs.~(\ref{eq:Friedmann})--(\ref{eq:matter_de}): new terms appear in the Friedmann (\ref{eq:Friedmann_tilde}) and acceleration (\ref{eq:acceleration_tilde}) equations and the right hand sides of Eqs.~(\ref{eq:matter_tilde_dm}) and (\ref{eq:matter_tilde_de}) are no longer one the opposite of the other.
Moreover the conservation of the total energy implies
\begin{equation}
	\dot{\tilde\rho}_{\rm int} + 3H (\tilde\rho_{\rm int} +\tilde{p}_{\rm int}) + \tilde{Q}_{dm} + \tilde{Q}_{de} = 0 \,,
	\label{eq:energy_constr}
\end{equation}
meaning that the new terms appearing in Eqs.~(\ref{eq:Friedmann_tilde})--(\ref{eq:matter_tilde_de}) are actually constrained.
Of course if an exact form for the Lagrangian (\ref{eq:Lag_tot}) would be provided, with all dependences on any degree of freedom well specified and all symmetries respected, then such a constraint would be automatically satisfied thanks to Noether's theorem.
Nevertheless in all the phenomenological models where the microscopical nature of the dark components is not postulated a priori, such a Lagrangian cannot be specified and the constraint (\ref{eq:energy_constr}) must be always considered together with Eqs.~(\ref{eq:Friedmann_tilde})--(\ref{eq:matter_tilde_de}).

The following question now arises naturally: do Eqs.~(\ref{eq:Friedmann_tilde})--(\ref{eq:energy_constr}) describe the same physics of Eqs.~(\ref{eq:Friedmann})--(\ref{eq:matter_de})?
Or to put it another way: are they two different representations of the same physical system?

In order to tackle such a question a reflection on the meaning of energy for two coupled physical systems is necessary.
Consider a closed physical system which can be divided into two sub-parts.
If these two sub-systems do not interact with each other, then one can always associate two energies which will be separately conserved.
However if the two sub-systems are interacting, i.e.~they are exchanging energy, then it is not possible to define two separately conserved energies, but only the total energy of the system is physically well defined.
The same situation applies to the cosmological dark sector since gravity allows only to probe its total energy and momentum \cite{Kunz:2007rk}.
In both Eqs.~(\ref{eq:Friedmann}) and (\ref{eq:Friedmann_tilde}) the only well defined energy density is thus the total energy density
\begin{equation}
	\rho_{\rm tot} = \rho_{dm} + \rho_{de} = \tilde\rho_{dm} + \tilde\rho_{de} + \tilde\rho_{\rm int} \,,
	\label{eq:tot_energy}
\end{equation}
which in fact is the one that sources the cosmological equations equalling $3H^2$.
Analogously in Eqs.~(\ref{eq:acceleration}) and (\ref{eq:acceleration_tilde}) the physically meaningful pressure is
\begin{equation}
	p_{\rm tot} = p_{dm} + p_{de} = \tilde{p}_{dm} + \tilde{p}_{de} + \tilde{p}_{\rm int} \,.
	\label{eq:tot_pressure}
\end{equation}
Thus any definition for the DE and DM energies and pressures is arbitrary, and any transformation between the tilde and the non-tilde quantities which leaves the total energy density and pressure invariant, will not alter the physics of the system.

Following this line of thought, the following completely general linear transformations, leaving the total energy density (\ref{eq:tot_energy}) and pressure (\ref{eq:tot_pressure}) invariant, can be defined between the tilde and non-tilde quantities
\begin{align}
	 {\rho}_{dm} &= \alpha\, \tilde\rho_{dm} + \beta\, \tilde\rho_{de} + \gamma\, \tilde\rho_{\rm int} \, \label{eq:transf_1} \\
	 {\rho}_{de} &= (1- \alpha) \tilde\rho_{dm} + (1-\beta) \tilde\rho_{de} + (1- \gamma) \tilde\rho_{\rm int} \,, \\
	 {p}_{dm} &= \alpha\, \tilde{p}_{dm} + \beta\, \tilde{p}_{de} + \gamma\, \tilde{p}_{\rm int} \,, \\
	 {p}_{de} &= (1- \alpha) \tilde{p}_{dm} + (1- \beta) \tilde{p}_{de} + (1- \gamma) \tilde{p}_{\rm int} \label{eq:transf_4} \,, 
\end{align}
where $\alpha$, $\beta$ and $\gamma$ are constant parameters.
Note that any transformation preserving Eqs.~(\ref{eq:tot_energy}) and (\ref{eq:tot_pressure}) could be employed for the considerations that follow.
The linear transformations (\ref{eq:transf_1})--(\ref{eq:transf_4}) have been chosen since they are quite simple and sufficiently general for our scopes.
Eqs.~(\ref{eq:Friedmann_tilde}) and (\ref{eq:acceleration_tilde}) clearly become Eqs.~(\ref{eq:Friedmann}) and (\ref{eq:acceleration}) after the transformations (\ref{eq:transf_1})--(\ref{eq:transf_4}) have been applied.
This corresponds to nothing but the invariance requirement for the total energy density and pressure.
It remains to understand if Eqs.~(\ref{eq:matter_tilde_dm}) and (\ref{eq:matter_tilde_de}) can be transformed into Eqs.~(\ref{eq:matter_dm}) and (\ref{eq:matter_de}).
Applying the transformation (\ref{eq:transf_1})--(\ref{eq:transf_4}), and taking into account the constraint (\ref{eq:energy_constr}), one finds
\begin{align}
	\dot{\rho}_{dm} + 3H \left( \rho_{dm} + p_{dm} \right) &= (\alpha- \gamma) \tilde{Q}_{dm} + (\beta-\gamma) \tilde{Q}_{de} \,, \nonumber \\
	\dot{\rho}_{de} + 3H \left( \rho_{de} + {p}_{de} \right) &= -(\alpha- \gamma) \tilde{Q}_{dm} - (\beta-\gamma) \tilde{Q}_{de} \,, \nonumber
\end{align}
which, defining
\begin{equation}
	Q = -(\alpha- \gamma) \tilde{Q}_{dm} - (\beta-\gamma) \tilde{Q}_{de} \,,
\end{equation}
coincide with Eqs.~(\ref{eq:matter_dm}) and (\ref{eq:matter_de}).

It seems thus that Eqs.~(\ref{eq:Friedmann_tilde})--(\ref{eq:energy_constr}) are indeed equivalent to Eqs.~(\ref{eq:Friedmann})--(\ref{eq:matter_de}), in the sense that they represent nothing but a different representation of the same physics.
Nevertheless in order to complete the equivalence one should show that also the EoS (\ref{eq:EoS_linear}) do not change and this is generally not the case.
To see this note that in principle from the ``bare'' Lagrangians $\mathcal{L}_{\rm dm}$ and $\mathcal{L}_{\rm de}$ one still derives the EoS $\tilde{p}_{dm} = w_{dm} \tilde\rho_{dm}$ and $\tilde{p}_{de} = w_{de} \tilde\rho_{de}$, since the interaction does not affect these relations.
This is no longer true once the redefinitions (\ref{eq:transf_1})--(\ref{eq:transf_4}) are applied and in fact one finds
\begin{equation}
	{p}_{dm} = w_{dm} \rho_{dm} + \pi_{dm} \quad\mbox{and}\quad {p}_{de} = w_{de} \rho_{de} + \pi_{de} \,,
	\label{eq:pressures_plus_dissipation}
\end{equation}
where
\begin{align}
	\pi_{dm} &= \beta \left( \tilde{p}_{de} - w_{dm} \tilde\rho_{de} \right) + \gamma \left( \tilde{p}_{\rm int} - w_{dm} \tilde\rho_{\rm int} \right) \,, \nonumber \\
	\pi_{de} &= (1-\alpha) \left( \tilde{p}_{dm} - w_{de} \tilde\rho_{dm} \right) + (1- \gamma) \left( \tilde{p}_{\rm int} - w_{de} \tilde\rho_{\rm int} \right) \,. \nonumber
\end{align}
For the sake of simplicity only linear EoS (\ref{eq:EoS_linear}) with $w_{de}$ and $w_{dm}$ constant have been considered.
However the same argument can be easily generalized to more general EoS as e.g.~$p_{dm}=f_{dm}(\rho_{dm})$ and $p_{de}=f_{de}(\rho_{de})$ with $f_{dm}$ and $f_{de}$ any two arbitrary functions.

Eqs.~(\ref{eq:pressures_plus_dissipation}) imply that any phenomenological model of DE interacting with DM should be defined specifying, in terms of other known quantities, the functions $\pi_{de}$ and $\pi_{dm}$ in addition to the energy exchange factor $Q$.
Fortunately one can always redefine these three functions in such a way that only two of them must be specified in the background equations.
In fact a simple and intuitive way to write the most general interacting DE equations at the background cosmological level is as follows
\begin{gather}
	3H^2 = \rho_{dm} + \rho_{de} \,, \label{eq:Friedmann_general} \\
	2\dot{H} + 3H^2 = -w_{dm} \rho_{dm} -w_{de} \rho_{de} - \pi \,, \label{eq:acceleration_general} \\
	\dot{\rho}_{dm} + 3H \rho_{dm} (1 + w_{dm}) = Q_{dm} \,, \label{eq:matter_dm_general} \\
	\dot{\rho}_{de} + 3H \rho_{de} (1 + w_{de}) = Q_{de} \,, \label{eq:matter_de_general}
\end{gather}
where the constraint
\begin{equation}
	3H\pi + Q_{dm} + Q_{de} = 0 \,,
	\label{eq:pi_constr}
\end{equation}
must be imposed and the definitions
\begin{gather}
	\pi = \pi_{dm} + \pi_{de} \,, \\
	Q_{dm} = -Q - 3H \pi_{dm} \,,\quad Q_{de} = Q - 3H \pi_{de} \,,
\end{gather}
have been applied.

According to Eq.~(\ref{eq:pi_constr}) $\pi$ measures the energy lost in the interaction between DE and DM and adsorbed by the gravitational field.
In other words it represents a {\it bulk dissipative pressure} characterizing a viscous interaction between the two dark fluids.
That this effect must be included can be intuitively understood realizing that the physical system under consideration is actually composed by three fluids: DM, DE and the gravitational field.
The most general situation, described by Eqs.~(\ref{eq:Friedmann_general})--(\ref{eq:pi_constr}), allows thus for the energy to be exchanged not only between DE and DM, but also with gravity.
For the most general models of DE interacting with DM one must thus specify two functions, rather than only the exchange factor $Q$ as in Eqs.~(\ref{eq:Friedmann})--(\ref{eq:EoS_linear}).
For example the functions $Q_{dm}$ and $Q_{de}$ can be provided, while $\pi$ is given by the constraint (\ref{eq:energy_constr}).
Clearly whenever $\pi=0$ the system reduces to the one described by Eqs.~(\ref{eq:Friedmann})--(\ref{eq:EoS_linear}), i.e.~to the standard literature approach.
In this simpler case DM and DE are effectively described by perfect fluids because no dissipation is present. 
However a sufficiently small bulk viscosity, which is the only dissipative effect allowed in a homogeneous and isotropic universe, might suppress structure formation at small scales, solving in this way some of the pathologies of standard cold DM \cite{Chimento:2003iea,Velten:2014xca,Komatsu:2015nga}.
The results presented here suggest that such viscosity might be due to the interaction between DM and DE.

\section{Covariant Equations}

It remains to understand if the same arguments apply at the fully covariant level from which the dynamics for the cosmological perturbations can be derived.
At the covariant level Eqs.~(\ref{eq:matter_dm}) and (\ref{eq:matter_de}) generalize to
\begin{equation}
	\nabla^\mu {T}_{\mu\nu}^{(dm)} = -{Q}_\nu \quad\mbox{and}\quad
    \nabla^\mu {T}_{\mu\nu}^{(de)} = {Q}_\nu \,,
    \label{eq:cov_matter_eqs}
\end{equation}
where ${T}_{\mu\nu}^{(dm)}$ and ${T}_{\mu\nu}^{(de)}$ are respectively the energy-momentum tensors of DM and DE and $Q_\mu$ is a four-vector dictating the amount of energy-momentum exchanged between the two dark components.
The coupling is introduced in order to preserve the conservation of the total energy momentum tensor,
\begin{equation}
	\nabla^\mu \left( {T}_{\mu\nu}^{(dm)} + {T}_{\mu\nu}^{(de)} \right) = 0 \,,
\end{equation}
so that the standard Einstein field equations
\begin{equation}
	G_{\mu\nu} = {T}_{\mu\nu}^{(dm)} + {T}_{\mu\nu}^{(de)} \,,
	\label{eq:einstein_eq}
\end{equation}
remain consistent with the Bianchi identity and can be employed to describe the dynamics of gravitation.
In any non-interacting cosmological model $Q_\mu$ vanishes identically and the two energy-momentum tensors are separately conserved.
However if a dark sector interaction is at work, then $Q_\mu$ is non-zero and it might potentially depend on all degrees of freedom of the dark sector.
For example, in a simple two scalar field toy model one would expect $Q_\mu$ to depend on both the scalar fields and possibly to be derived from the interacting potential energy coupling the two fields at the Lagrangian level.
Nevertheless in the absence of a microscopical description of the dark components, $Q_\mu$ is phenomenologically assumed to depend on macroscopic quantities such as the energy density, pressure and local velocity of a fluid.
A common expression is for example $Q_\mu = Q\, u_\mu$, where $u_\mu$ can either be the fluid velocity of DM, DE or a combination of both, although for these specific models one must be careful that instabilities do not arise at the cosmological perturbations level \cite{Li:2013bya,Valiviita:2008iv,Majerotto:2009np,He:2008si,Li:2014eha}.

To understand whether Eqs.~(\ref{eq:cov_matter_eqs})--(\ref{eq:einstein_eq}) represent the most general way to couple DE and DM at the covariant level, the Lagrangian (\ref{eq:Lag_tot}) will again be employed to derive interacting equations from a supposedly unknown fundamental level.
The variation with respect to the metric and the dark sector variables produces the general equations
\begin{gather}
	G_{\mu\nu} = \tilde{T}_{\mu\nu}^{(dm)} + \tilde{T}_{\mu\nu}^{(de)} + \tilde{T}_{\mu\nu}^{\rm (int)} \,, \label{eq:einstain_ex} \\
	\nabla^\mu \tilde{T}_{\mu\nu}^{(dm)} = \tilde{Q}_\nu^{(dm)} \quad\mbox{and}\quad
    \nabla^\mu \tilde{T}_{\mu\nu}^{(de)} = \tilde{Q}_\nu^{(de)} \label{eq:matter_ex} \,,
\end{gather}
where $\tilde{T}_{\mu\nu}^{\rm (int)}$, $\tilde{Q}_\nu^{(dm)}$ and $\tilde{Q}_\nu^{(de)}$ are associated with the variation of $\mathcal{L}_{\rm int}$ with respect to $g_{\mu\nu}$, $\psi_{dm}$ and $\psi_{de}$, respectively.
The conservation of the total energy-momentum yields the constraint
\begin{equation}
	\nabla^\mu \tilde{T}_{\mu\nu}^{\rm (int)} + \tilde{Q}_\nu^{(dm)} + \tilde{Q}_\nu^{(de)} = 0 \,,
	\label{eq:cov_energy_constr}
\end{equation}
which would be identically satisfied the complete expression of the Lagrangian had been known.

Generalizing the arguments outlined above for the background equations to a fully relativistic situation, the only physically meaningful energy-momentum tensor in the system can be identified with the total energy-momentum tensor
\begin{equation}
	T_{\mu\nu}^{\rm (tot)} = {T}_{\mu\nu}^{(dm)} + {T}_{\mu\nu}^{(de)} = \tilde{T}_{\mu\nu}^{(dm)} + \tilde{T}_{\mu\nu}^{(de)} + \tilde{T}_{\mu\nu}^{\rm (int)} \,.
\end{equation}
The definitions of ${T}_{\mu\nu}^{(dm)}$ and ${T}_{\mu\nu}^{(de)}$ are in fact arbitrary and the linear transformations
\begin{align}
	{T}_{\mu\nu}^{(dm)} &= \alpha \tilde{T}_{\mu\nu}^{(dm)} + \beta \tilde{T}_{\mu\nu}^{(de)} + \gamma \tilde{T}_{\mu\nu}^{\rm (int)} \,, \label{eq:cov_transf_1} \\
	{T}_{\mu\nu}^{(de)} &= (1-\alpha) \tilde{T}_{\mu\nu}^{(dm)} + (1-\beta) \tilde{T}_{\mu\nu}^{(de)} + (1-\gamma) \tilde{T}_{\mu\nu}^{\rm (int)} \,, \label{eq:cov_transf_2}
\end{align}
with $\alpha$, $\beta$ and $\gamma$ constants, do not alter the physical content of the system since they leave the total energy-momentum tensor invariant.
Applying these transformations to Eqs.~(\ref{eq:einstain_ex}) and (\ref{eq:matter_ex}) one immediately recovers the Einstein equations (\ref{eq:einstein_eq}), while the matter equations (\ref{eq:cov_matter_eqs}) are obtained, with the use of the constraint (\ref{eq:cov_energy_constr}), as
\begin{align}
\nabla^\mu {T}_{\mu\nu}^{(dm)} &= (\alpha- \gamma) \tilde{Q}_\nu^{(dm)} + (\beta-\gamma) \tilde{Q}_\nu^{(de)} = -{Q}_\nu \,,\label{eq:matter_im_1}\\
\nabla^\mu {T}_{\mu\nu}^{(de)} &= -(\alpha - \gamma) \tilde{Q}_\nu^{(de)} -(\beta- \gamma) \tilde{Q}_\nu^{(de)} = {Q}_\nu \,.\label{eq:matter_im_2}
\end{align}
These coincide with Eqs.~(\ref{eq:cov_matter_eqs})--(\ref{eq:einstein_eq}) which thus describe the most general way of coupling DM to DE at the covariant level.
However one must keep in mind that the definitions of ${T}_{\mu\nu}^{(dm)}$ and ${T}_{\mu\nu}^{(de)}$ used in this case generally depend on each other's variables since they implicitly contain the interacting term $\tilde{T}_{\mu\nu}^{\rm (int)}$, as it is clear from the transformations (\ref{eq:cov_transf_1}) and (\ref{eq:cov_transf_2}).
Hence requiring the DE and DM energy-momentum tensors to be not modified by the dark sector interaction, equals to implicitly assume that $\tilde{T}_{\mu\nu}^{\rm (int)}=0$, which is generally not the case.
To give an example, if ${T}_{\mu\nu}^{(de)}$ is taken to be of the fluid type, in agreement with the cosmological principle, then not only the energy density and pressure associated with it will supposedly depend on the DM degrees of freedom, but also its four-velocity cannot be straightforwardly identified with the DE four-velocity because of the interaction with DM.
At the perturbation level thus the three-velocity derived from ${T}_{\mu\nu}^{(de)}$ will generally not coincide with the DE velocity, being it affected by the DM coupling.
How to define a meaningful covariant coupling for phenomenological models of DE interacting with DM, i.e.~when their microscopical nature is unknown, is still an open issue \cite{Faraoni:2014vra}, partly due to the difficulty of describing all possible dissipative effects.
Since only at the Lagrangian level the full dynamics can be consistently specified, a possible solution might be the construction of an effective action describing the interaction.
However, although some steps forward have been recently made with different fluid Lagrangian formalisms \cite{Pourtsidou:2013nha,Skordis:2015yra,Ballesteros:2013nwa,Boehmer:2015kta,Boehmer:2015sha,Tomi-Manos,Gleyzes:2015pma}, such an effective approach has still to be fully developed.

\section{Examples}

In order to better show how the results obtained in this work can apply to specific models of dark sector interactions, in what follows two examples will be provided.
The first one consists in a straightforward generalization of already well-known phenomenological models of DE coupled to DM, while the second one presents a recently proposed formulation for a scalar field interacting with DM, where one can explicitly see a direct application of the results obtained in the previous sections.
The aim of this section is only to emphasize and briefly explain the new possible features arising from the models considered in the following.
An in depth analysis of their cosmological dynamics at both background and perturbation level will be left for future studies or, when possible, reference to current literature will be provided.

\subsection{Phenomenological models}

As already mentioned, the most popular phenomenological couplings between DE and DM are defined, at least at the background level, by $Q \propto \rho$ and $Q \propto H\rho$, where $\rho$ stands for the energy density of DM, DE or a combination of the two.
To study the cosmological dynamics of these models Eqs.~(\ref{eq:matter_dm})--(\ref{eq:matter_de}) plus Eqs.~(\ref{eq:Friedmann})--(\ref{eq:acceleration}) are commonly employed, implicitly assuming that the two right hand sides are one the opposite of the other.
However, as it has been shown above, the most general background equations for such models are actually Eqs.~(\ref{eq:Friedmann_general})--(\ref{eq:matter_de_general}), together with the constraint (\ref{eq:pi_constr}).
Among these four equations only three are actually independent due to the total conservation of energy and thus one can choose to discard one of them, say Eq.~(\ref{eq:acceleration_general}), when investigating the dynamics of the system.

Given these considerations, a simple extension of the phenomenological model with interaction $Q \propto \rho$ can be described by the following equations
\begin{gather}
	3H^2 = \rho_{dm} + \rho_{de} \,, \label{eq:Friedmann_example} \\
	\dot{\rho}_{dm} + 3H \rho_{dm} (1 + w_{dm}) = \alpha\, \rho \,, \label{eq:matter_dm_example} \\
	\dot{\rho}_{de} + 3H \rho_{de} (1 + w_{de}) = \beta\, \rho \,, \label{eq:matter_de_example}
\end{gather}
where $\alpha$, $\beta$ are two constants and $\rho$ stands again for a general combination of $\rho_{dm}$ and $\rho_{de}$, but usually is taken to coincide with one of them for the sake of simplicity.
Note that now the right hand sides of Eqs.~(\ref{eq:matter_dm_example})--(\ref{eq:matter_de_example}) are not the opposites of each other, reflecting the fact that the bulk dissipative pressure $\pi$ is actually non zero.
In fact from the constraint (\ref{eq:pi_constr}) one immediately obtains
\begin{equation}
	3H\pi = - (\alpha + \beta)\, \rho \,,
\end{equation}
which clearly states that $\pi \neq 0$ unless $\alpha = - \beta$, i.e.~unless the commonly assumed case $Q_{dm} = -Q_{de}$ is considered.

At the fully covariant level the coupling $Q\propto \rho$ is usually generalized to $Q_\mu \propto \rho\, u_\mu$, where $u_\mu$ is the fluid velocity of DM or DE.
This procedure, although not unique, allows for a consistent investigation of the cosmological perturbations and instabilities have been found in the cases where $\rho$ coincides with $\rho_{dm}$ or $\rho_{de}$ \cite{Valiviita:2008iv}, while more complex combinations for $\rho$ seem to be viable \cite{Li:2013bya}.
The simplest extension of these models, in agreement with Eqs.~(\ref{eq:Friedmann_example})--(\ref{eq:matter_de_example}), consists thus in taking $Q_\mu^{(dm)} = \alpha \rho u_\mu$ and $Q_\mu^{(de)} = \beta \rho u_\mu$.
The analysis of cosmological perturbations for these new models is expected to be very similar to the analysis already performed for the well-known models where $\alpha = -\beta$. 
Nevertheless new results could be found due to the differences introduced by the new effects; for example the simplest models where $\rho$ is identified with $\rho_{dm}$ or $\rho_{de}$ might be actually stable.
Note however that from the considerations outlined in the previous sections one should also expect the effects of the interaction to implicitly appear in the energy-momentum tensors ${T}_{\mu\nu}^{(dm)}$ and ${T}_{\mu\nu}^{(de)}$.
In other words each of the energy-momentum tensor, or at least one of them, should depend on the degrees of freedom of the other.
This fact will be more explicit in the following example where all the degrees of freedom of DE and DM are specified, at least within an effective approach, from a Lagrangian formulation.

\subsection{Scalar-Fluid theories} 

In this second example the recently proposed framework of Scalar-Fluid theories \cite{Boehmer:2015kta,Boehmer:2015sha,Tomi-Manos} will be used to emphasize some of the aspects discussed in the main body of this work.
Following \cite{Boehmer:2015kta,Boehmer:2015sha,Tomi-Manos} the Lagrangian of these theories can be written as
\begin{equation}
	\mathcal{L}_{SF} = \mathcal{L}_{\rm EH} + \mathcal{L}_\phi + \mathcal{L}_M + \mathcal{L}_{\rm int} \,,
	\label{eq:SF_Laggrangain}
\end{equation}
where $\mathcal{L}_{\rm EH}$ is the standard Einstein-Hilbert Lagrangian for general relativity and $\mathcal{L}_\phi$ is a canonical scalar field Lagrangian with potential $V(\phi)$.
The Lagrangian of DM is $\mathcal{L}_M$, which in this formalism is assumed to be described by an effective matter fluid, while $\mathcal{L}_{\rm int}$ characterizes the interacting term between DM and the scalar field representing DE.
The details of the Lagrangian (\ref{eq:SF_Laggrangain}) will not be exposed here, but the reader is referred to \cite{Boehmer:2015kta,Boehmer:2015sha,Tomi-Manos} for more information.

Note how the Lagrangian (\ref{eq:SF_Laggrangain}) is exactly in the form of Eq.~(\ref{eq:Lag_tot}) and in fact the interacting term $\mathcal{L}_{\rm int}$ indeed contains both DE (the scalar field) and DM (the effective fluid variables) degrees of freedom.
The direct variation of (\ref{eq:SF_Laggrangain}) will then give rise to equations in the form (\ref{eq:einstain_ex})--(\ref{eq:matter_ex}), as explicitly shown in \cite{Boehmer:2015kta,Boehmer:2015sha}.
However with a suitable redefinition of the energy-momentum tensors, corresponding to a specific transformation (\ref{eq:cov_transf_1})--(\ref{eq:cov_transf_2}), the same equations can be written in the form (\ref{eq:cov_matter_eqs})--(\ref{eq:einstein_eq}) corresponding to the formulation used in \cite{Tomi-Manos}.
In what follows only this latter formulation will be considered since it better compares to the standard equations of DE interacting with DM.

The exchange vector for these models reads\footnote{See Eqs.~(2.29) in \cite{Tomi-Manos}. Here only the so-called derivative couplings will be considered since they better highlight the issues discussed in what follows.}
\begin{equation}
 Q_\nu  =   -n^2\frac{\partial f(n,\phi)}{\partial n}\nabla_\lambda u_{(dm)}^\lambda\partial_\nu\phi  \,.
\end{equation}
while the DM pressure is given by
\begin{equation}
	p_{dm}   = w_{dm} \rho_{dm} -n^2\frac{\partial f(n,\phi)}{\partial n} u_{(dm)}^\lambda\partial_\lambda\phi \,.
	\label{eq:pressure_dm_SF}
\end{equation}
Here $n$ is the particle number density of DM, $\phi$ is the scalar field representing DE, $f$ is a general function of $n$ and $\phi$, $u_{(dm)}^\lambda$ is the four-velocity of DM and $\nabla_\lambda$ is the covariant derivative with respect to $g_{\mu\nu}$.
The DE energy-momentum tensor is provided by its canonical expression and thus no dependence on DM is present.
On the other hand the DM energy-momentum tensor is of the fluid type with four-velocity $u_{(dm)}^\lambda$, energy density $\rho_{dm}$ and pressure given by Eq.~(\ref{eq:pressure_dm_SF}).
Note that in this latter energy-momentum tensor a dependence on the DE degrees of freedom is indeed present since the scalar field appears in the DM pressure, as explicitly shown in Eq.~(\ref{eq:pressure_dm_SF}).
This situation formally corresponds to the transformations (\ref{eq:cov_transf_1})--(\ref{eq:cov_transf_2}) with $\alpha = \gamma = 1$ and $\beta = 0$, i.e.~when the DE energy-momentum tensor retains its form and all the interacting terms are adsorbed by the DM side.

The background equations of these Scalar-Fluid models can be written as
\begin{gather}
	3H^2 = \rho_{dm} + \rho_{de} \,, \\
	2\dot{H} + 3H^2 = -w_{dm} \rho_{dm} -w_{de} \rho_{de} +n^2\frac{\partial f}{\partial n} \dot\phi \,, \\
	\dot{\rho}_{dm} + 3H \rho_{dm} (1 + w_{dm}) = Q_0 +n^2\frac{\partial f}{\partial n} \dot\phi \,,  \\
	\dot{\rho}_{de} + 3H \rho_{de} (1 + w_{de}) = -Q_0 \,,
\end{gather}
where
\begin{equation}
	Q_0 = -3Hn^2 \frac{\partial f}{\partial n} \dot\phi \,.
\end{equation}
Note how these equations can be recast into Eqs.~(\ref{eq:Friedmann_general})--(\ref{eq:matter_de_general}) once the identifications
\begin{equation}
	\pi = -n^2\frac{\partial f}{\partial n} \dot\phi \,,\quad Q_{dm} = Q_0 +n^2\frac{\partial f}{\partial n} \dot\phi \,,\quad Q_{de} = -Q_0 \,,
\end{equation}
have been made.
Note also that the constraint (\ref{eq:pi_constr}) is automatically satisfied as expected since the equations have been derived from a suitable Lagrangian.
This example explicitly shows how models with $Q_{dm} \neq -Q_{de}$ can actually appear when an interaction in the dark sector is switched on.
Moreover it suggests that whenever the coupling between DM and DE is defined at the Lagrangian level, such situation will generally define the background equations, unless the interaction in the Lagrangian is chosen such that $\pi=0$ in the resulting equations.
In the case of Scalar-Fluid theories considered here the bulk dissipative pressure $\pi$ is indeed non-vanishing and in general depends on both DE (through $\phi$) and DM (through $n$) degrees of freedom.

Finally the fact that the cosmological equations for Scalar-Fluid theories are derived from a properly constructed Lagrangian, implies that no ambiguities arise at the perturbation level.
In fact, in contrast to the situation for more phenomenological models, the fully covariant energy-momentum tensors of DE and DM are always well-defined in terms of all degrees of freedom.
Nevertheless the dependence of the DM energy-momentum tensor upon the scalar field results in a more complicated dynamics for the cosmological perturbations as properly exposed in \cite{Tomi-Manos}.
As a consequence new terms due to the interaction appear in the perturbation equations and one must be careful that they do not introduce new instabilities, especially due to possible non-linearities.
A thorough discussion on the dynamics of Scalar-Fluid theories at the perturbation level is complicated and a detailed analysis is clearly outside the scope of the present work.
The interested reader can refer to \cite{Tomi-Manos} where these models have been studied in depth and the differences with respect to more restrictive models have been extensively discussed.

\section{Conclusion}

In conclusion the analysis presented in this work shows that a phenomenological interaction in the dark sector is likely to affect the cosmological dynamics not only modifying the matter conservation equations, as considered in the literature so far, but also introducing new terms in the gravitational field equations.
For example in the background cosmological equations the only possible modification due to the coupling of DE to DM is the addition of a bulk dissipative pressure, describing viscous phenomena between the two fluids, in the acceleration equation.
Interestingly this kind of effects might be made responsible for the suppression of structure formation at small scales, helping in this way to reconcile theory and observations.
At the generally covariant level instead, the discussion presented in this work highlights the problems in defining physically relevant quantities, e.g.~the velocity, associated with the separate fluids.
These issues directly affect the cosmological perturbations signaling the difficulties, already emerged several times in the literature on the subject, in properly studying their dynamics when an interaction in the dark sector is not provided by some well-defined Lagrangian formulation.
The overall result of the present work is thus to formally connect all these problems regarding phenomenological models of DE coupled to DM to the lack of a proper fundamental approach.

\bibliography{bibfile}

\end{document}